\documentclass[11pt]{article}
\usepackage{amsmath, mathtools, amssymb, amsthm, amsfonts}
\usepackage[numbers,comma,sort&compress]{natbib}
\usepackage{graphicx}
\usepackage{color}
\usepackage{caption}
\usepackage{subcaption}
\usepackage{float}
\usepackage{placeins}
\usepackage{listings}
\usepackage{multicol}
\usepackage{bicaption}
\usepackage{array}
\usepackage{cite}

\usepackage{lipsum}
\usepackage{graphicx}
\usepackage{url}
\usepackage{authblk}

\date{}
\title{\bf Criteria for the CloudSim Environment}
\author[1]{ Arezoo  Khatibi\thanks{Arezoo  Khatibi, correspond email: \protect\url{arezoo.khatibi@grad.kashanu.ac.ir}, Faculty of Computer Science,  University of Kashan,  BLVD Ghotb Ravandi 6 kilometers, Kashan Iran.}}
\author[2]{Omid Khatibi}
\affil[1]{Faculty of Computer Science, University of Kashan, Kashan, Iran}
\affil[2]{Faculty of Mathematics, University of Vienna,Vienna, Austria}

\begin{document}
\maketitle

\begin{abstract}
CPU is undoubtedly the most important resource of the computer system. Recent advances in software and system architecture have increased  processing complexity, as  computing is now distributed and parallel. CloudSim represents the complexity of an application in terms of its computational requirements. CloudSim [9] is a complete solution for simulating Cloud Computing environments and building test beds for provisioning algorithms. This paper is analyzes and evaluates the performance of cloud environment modeling using CloudSim. We describe the CloudSim architecture and then investigate the new models and techniques in CloudSim.
\end{abstract}
\section{Introduction}
 The national institute of technology (NIST) has formulated a widely accepted definition that characterizes important aspects of cloud computing: Cloud computing is a model for enabling ubiquitous, convenient, on-demand network access to a shared pool of configurable computing resources (e.g. networks, servers, storage, applications, and services) that can be rapidly provisioned and released with minimal management effort or service provider interaction.  How Websites relate to  memory use, processor, and small hardware depends on the type of server the website uses  and this can be problematic. As  data and website users   are increasing, perhaps a server cannot respond to this traffic but cloud computing can because it uses several servers for responsibility. That means a server always exists to carry additional load or several servers are considered for various works. In  the near future, we will not need OS, infrastructure, memory, storage, CPU or even application in our home and instead will meet these needs  through the sharing of cloud service providers by connecting to the Internet. In  the cloud, infrastructure, operating systems, hardware, and software requirements are provided as a service. There are three service models for cloud computing: Cloud infrastructure as a service (IaaS) cloud platform as a service (PaaS)  and cloud software as a service (SaaS). Three things must be provided for cloud computing services which are thin client,  grid computing, and  utility computing. Today, cloud computing is one step ahead of  utility computing and the provision of resources is on demand. In our CloudSim examples, we can change the virtual machine (VM) allocation policy, provision of resources, power consumption. For example, we can change the power folder  for low power consumption and test it.
 
  The following example are fetched from CloudSim examples folders [14].
Basic Examples: In this set,  there are eight  possible cloud examples wich are determined by a single host with one or two data centers  running one or two cloudlets on them, and using  with or without using network typology. In one of these examples, cloud Scalability is simulated and  the last two examples represet how to pause and resume  a cloud simulation, such as Data center broker. Network Examples: This set consists of four examples that illustrating how to create one or two data centers with one host and network topology which executes one or two user cloudlets. Power Examples: In these examples, simulation of a heterogeneous non-power aware data center or heterogeneous power aware data center is shown. In one  example, no optimization was performed on VM assignment. In  other examples, there are various policies for assigning  virtual machine (VM) such as  median absolute deviation (MAD), VM allocation or the local regression (LR), VM allocation policy or minimum migration time (MMT), VM selection policy choosing the most suitable VM, and so on.
\section{Scheduling in cloud }
 CloudSim is used to simulate the cloud and the cloudlet represents the applications. Cloud brokers assign  cloudlets to  virtual machines (VMs) and the virtual machine manager decides on which VM will host based on the VM allocation policy. The VMs  then start executing cloudlets and scheduling algorithms are introduced. We use Cloud to make better use of resources. The discussion of  scheduling algorithms is more common in parallel computer systems in which a task is divided into several subtasks and each subtask is run by one processor. When VM performs a task, it seems to have performed that task in parallel. For this purpose, it uses fare share scheduling or capacity scheduling. In CloudSim, there is a class that simulates VM and a host can simultaneously multiply initial  VMs. In this regard, the time  and space shared policies are used to allocate the cores and each VM is in relation to a component that holds the VM specification, such as memory and  scheduling algorithm. Also, workfloware  tasks  are composed of several sub-tasks.
\section{Cloud computing }
Cloud has emerged as an alternative to clusters and grids and many modern cloud services are provided using Internet Data Centers (IDCs) such as the Google search engine. Cloud computing is considered to be energy and ecological efficient. In Cloud computing, it is important to know how much energy a specific service or  VM consumes. Today, the internet involves  more than bandwidth resources. Computing and storage resources are shared through Cloud Computing offering virtual machines  over infrastructure services, application programming interfaces (APIs)  and support  through  service platformse, and Web-based applications to end users through software services.
\section{CloudSim architecture }
\begin{figure}[htbp]
 \centering
        \includegraphics{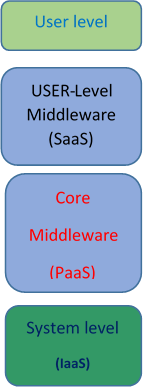}
        \caption{CloudSim architecture}
        \label{Graph A}
        \end{figure}
The architecture of the CloudSim includes the following layers:  The system level in which cloud resources are located (IaaS), core middleware (PaaS) is higher level that includes VM management and deployment, SLA management and monitoring and QoS negotiation. Above this level, there is a User-Level middleware (SaaS) that includes environments and tools, workflows and Cloud programming, and the highest level is User Level that includes Cloud applications.
        
      {   
       \begin{figure}[htbp]
 \centering
        \includegraphics{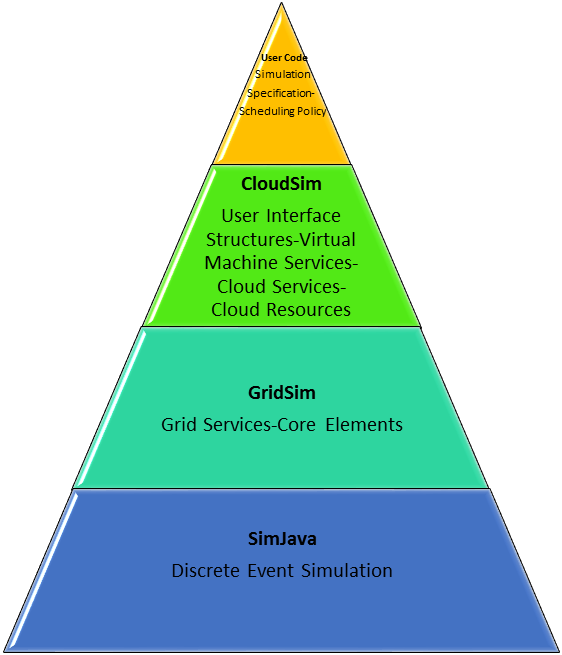}
        \caption{Layereering of cloud architecture}
        \label{Graph B}
        \end{figure}  }       
 The second figure shows the layered implementation of the CloudSim software framework. The SimJava layer implements system components (such as services, host, data center, broker, virtual machines) the simulates communication between these components, and manages simulation clock. GridSim [9] supports high-level software components for modeling multiple Grid infrastructures, including networks and Grid components (such as  resources, data sets, workload traces, and information services) and SimJava avoids reimplementation of event handling and message passing among components.
\begin{figure}[htbp]
 \centering
 \includegraphics{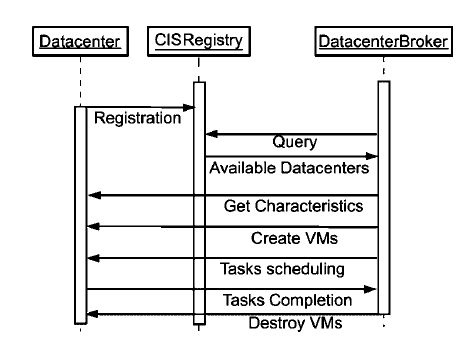}
        \caption{Simulation data flow in CloudSim}
        \label{Graph C}
        \end{figure}\newline 
        
 The third figure shows the communication between core CloudSim entities. At the beginning of a simulation, each Data Center entity registers with the CIS (Cloud Information Service) Registry. Next, the Data Center brokers try to obtain the list of Cloud providers.  The Data Center broker  then deploys the application [11], and looks for the best VM to run the application.
 
\section{Our experience}
Initially,we implemented the program using one cloudlet for simplicity then ran the prgram with more cloudlets. Each VM has a virtual CPU and in CloudSim, the virtual CPU is called Processing Element (PE). The VM can have one PE or more which simulates the original multi-core CPUs. Here, we will describe the run of our program and the steps performed. First , initialize the CloudSim package, it should be called before creating any entities (number of grid users, trace events, calendar, and initializing the CloudSim library). Second, create Data Centers as they are the resource providers in CloudSim and we need at least one of them to run a CloudSim simulation. when creating a Data Center, there are several considerations to follow: We must create a list to store one or more Machines and a Machine contains one or more Machines and a Machine contains one or more PEs or CPUs/cores, thus we must create a list to store the PEs before creating a Machine. So, we create a list for four PEs (four are required) for a quad core Machine and create another list for a dual core Machine. Next, Create  a Host id with the list of PEs and add them to the list of Machines. Then, Create a Data Center Charcteristics object to stores data center properties of architecture, OS, list of Machines and allocation policy( time- or space-shared, time zone, and  price of (\text{G}\textdollar/per  unit of time). The last consideration for creating  Data Center  is to  create a power Data Center Object.  The third step of the program is  to create a broker followed by the fourth step to creat VMs and Cloudlets to send  to the broker. The fifth step is to Start the simulation and the final step is to print  the results of the simulation. 

There are several steps to creating a VM. First, create a container to store  the VMs and this list is passed to the broker later. Second, Specify VM Parameters as image size, VM memory (MB), number of CPUs and VMM name. Next, Create a VMs array. We use these values in our program: VM image size of  10000, VM memory (MB) of 512, and  one CPU. We set  the cloudlet length  to 1000 and we set the cloudlet file size to 300. Our operating system is Windows and our system architecture is x86. We will calculate two values, the completion rate and the average execution time in our program. The completion rate, or percentage of work completed is equal to the total number of cloudlets  divided by the number of successful cloudlets. The average execution time of each cloudlet is determined by the number of cloudlets divided by the execution time. We show the number of VMs illustrated by the red line and growth of the average execution time represented by the blue columns in Figure 4  The result of the first run of our CloudSim program can be in table 1.

\begin{figure}[htbp]
 \centering
 \includegraphics{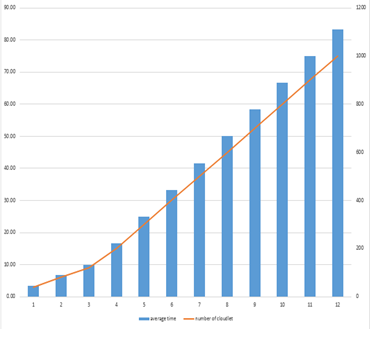}
        \caption{The number of VMs  and growth of the average execution time.}
        \label{Graph D}
        \end{figure}
        \begin{minipage}{\linewidth}
 \begin{center}    
 \captionof{table}{The results of the first run of our CloudSim program.}\label{tab:Show the result of the first run of our CloudSim program}  
\begin{tabular}{ c c c c c c c }
Cloudlet ID & Status & Data center ID & VM ID & Time & Start Time & Finish Time \\\hline
4 & SUCCESS & 2 & 4 & 3 & 0.2 & 3.2  \\\hline
16 & SUCCESS & 2 & 4 & 3 & 0.2 & 3.2  \\\hline
28 & SUCCESS & 2 & 4 & 3 & 0.2 & 3.2  \\\hline
5 & SUCCESS & 2 & 5 & 3 & 0.2 & 3.2   \\\hline
17 & SUCCESS & 2 & 5 & 3 & 0.2 & 3.2 \\\hline
29 & SUCCESS & 2 & 5 & 3 & 0.2 & 3.2 \\\hline
6 & SUCCESS & 3 & 6 & 3 & 0.2 & 3.2 \\\hline
18 & SUCCESS & 3 & 6 & 3 & 0.2 & 3.2 \\\hline
30 & SUCCESS & 3 & 6 & 3 & 0.2 & 3.2 \\\hline
7 & SUCCESS & 3 & 7 & 3 & 0.2 & 3.2 \\\hline
19 & SUCCESS & 3 & 7 & 3 & 0.2 & 3.2 \\\hline
31 & SUCCESS & 3 & 7 & 3 & 0.2 & 3.2   \\\hline
8 & SUCCESS & 3 & 8 & 3 & 0.2 & 3.2 \\\hline
20 & SUCCESS & 3 & 8 & 3 & 0.2 & 3.2 \\\hline
32 & SUCCESS & 3 & 8& 3 & 0.2 & 3.2 \\\hline
    10&                SUCCESS&        3&                      10&              3 &        0.2&            3.2 \\\hline
22&                SUCCESS&        3&                      10&              3&        0.2&            3.2 \\\hline
    34&                SUCCESS&        3&                      10&              3&        0.2&            3.2 \\\hline
    9&                  SUCCESS&        3&                       9&               3&        0.2&            3.2 \\\hline
21&                SUCCESS&        3&                       9&               3&        0.2&            3.2 \\\hline
33&                SUCCESS&        3&                       9&               3&        0.2&            3.2  \\\hline
11&                SUCCESS&        3&                      11&              3&        0.2&            3.2  \\\hline
23&                SUCCESS&        3&                      11&              3&        0.2&            3.2  \\\hline
35&                SUCCESS&        3&                      11&              3&        0.2&            3.2 \\\hline
0&                  SUCCESS&        2&                       0&               4&        0.2&            4.2 \\\hline
12&                     SUCCESS&        2&                           0&            4&            0.2&             4.2 \\\hline
    24&                  SUCCESS&        2&                           0&            4&            0.2&            4.2 \\\hline
    36&                  SUCCESS&        2&                           0&            4&            0.2&            4.2 \\\hline
1&                    SUCCESS&        2&                           1&            4&            0.2&            4.2 \\\hline
13&                 SUCCESS&        2&                            1&            4&            0.2&            4.2 \\\hline
    25&                 SUCCESS&        2&                            1&            4&            0.2&            4.2 \\\hline
37&                 SUCCESS&        2&                            1&            4&            0.2&            4.2 \\\hline
2&                   SUCCESS&        2&                            2&            4&            0.2&            4.2 \\\hline
14&                 SUCCESS&        2&                            2&            4&            0.2&            4.2 \\\hline
    26&                 SUCCESS&        2&                            2&            4&            0.2&            4.2 \\\hline
38&                 SUCCESS&        2&                            2&            4&            0.2&            4.2 \\\hline
    3&                   SUCCESS&        2&                            3&            4&            0.2&            4.2 \\\hline  
 \end{tabular}
 \end{center}
\end{minipage}

\begin{minipage}{\linewidth}
\begin{center} 
 \begin{tabular}{ c c c c c c c } 
 Cloudlet ID & Status & Data center ID & VM ID & Time & Start Time & Finish Time \\\hline
15&                 SUCCESS&        2&                            3&            4&            0.2&            4.2 \\\hline
    27&                 SUCCESS&        2&                            3           & 4&            0.2&            4.2 \\\hline
39&                 SUCCESS&        2&                            3&            4&            0.2&            4.2 \\\hline
\end{tabular}
\end{center}
\end{minipage}
\\
 Completion rate: 1\\
 Average execution time: 3.398000000000001 
 \section{Analysis}
 The output of this program compares the implementation time of various parallel applications (named cloudlet) creating 40 cloudlets ( with two or three Data Centers) and creating 20 VMs (1 to 20 for the number of virtual machines. We obtained this result where time with two Data Centers and four five VMs is three, the time is the same when the cloudlets run with three Data Centers and six to eleven VMs is the same, and  when we have two Data Centers with zero to three VMs  the time is the same and  equal to four. We can change the cost of  processing  use  memory use,  storage use  BW (bandwidth) use in our program. We quintuplicate these costs and obtain the same results. We run our program with 1000000 VMs and 160 cloudlets and obtain these results:\\
   
\begin{minipage}{\linewidth}
        \begin{center}
        \captionof{table}{The result of the second run of our CloudSim program.}\label{tab:Show the result of the first run of our CloudSim program}  
          \begin{tabular}{ c c c c c c c }
Cloudlet ID & Status & Data center ID & VM ID & Time & Start Time & Finish Time \\\hline
    4&               SUCCESS&        2&                              4&            12.99&        0.2&            13.19\\\hline
    16&             SUCCESS&        2&                              4&            12.99&        0.2&            13.19\\\hline
    28&             SUCCESS&        2&                              4&            12.99&        0.2&            13.19\\\hline
40&             SUCCESS&        2&                              4&            12.99&        0.2&            13.19\\\hline
    52&             SUCCESS&        2&                              4&            12.99&        0.2&            13.19\\\hline
64&             SUCCESS&        2&                              4&            12.99&        0.2&            13.19\\\hline
76&             SUCCESS&        2&                              4&            12.99&        0.2&            13.19\\\hline
    88&             SUCCESS&        2&                             4&            12.99&        0.2&            13.19\\\hline
 100&           SUCCESS&        2&                             4&            12.99&        0.2&            13.19\\\hline   
 112&           SUCCESS&        2&                             4&            12.99&        0.2&            13.19 \\\hline
  124&           SUCCESS&        2&                             4&            12.99&        0.2&            13.19\\\hline
      136&           SUCCESS&        2&                             4&            12.99&        0.2&            13.19\\\hline
     148&           SUCCESS&        2&                             4&            12.99&        0.2&            13.19 \\\hline
      5&               SUCCESS&        2&                              5&            12.99&        0.2&            13.19\\\hline
     17&             SUCCESS&        2&                             5&            12.99&        0.2&            13.19 \\\hline
 29&             SUCCESS&        2&                             5&            12.99&        0.2&            13.19 \\\hline
  41&             SUCCESS&        2&                             5&            12.99&        0.2&            13.19\\\hline
  53&             SUCCESS&        2&                             5&            12.99&        0.2&            13.19\\\hline
      65&             SUCCESS&        2&                             5&            12.99&        0.2&            13.19\\\hline
  77&              SUCCESS&        2&                            5&            12.99&        0.2&            13.19\\\hline
      89&              SUCCESS&        2&                            5&            12.99&        0.2&            13.19\\\hline
  101&            SUCCESS&        2&                            5&            12.99&        0.2&            13.19\\\hline
  113&            SUCCESS&        2&                            5&            12.99&        0.2&            13.19\\\hline
  125&            SUCCESS&        2&                            5&            12.99&        0.2&            13.19\\\hline
  137&            SUCCESS&        2&                            5&            12.99&        0.2&            13.19\\\hline
      149&            SUCCESS&        2&                            5&            12.99&        0.2&            13.19\\\hline
      6&                SUCCESS&        3&                            6&            12.99&        0.2&            13.19\\\hline  
     18&             SUCCESS&        3&                             6&            12.99&        0.2&            13.19\\\hline
     30&             SUCCESS&        3&                              6&            12.99&        0.2&            13.19\\\hline
 42&             SUCCESS&        3&                              6&            12.99&        0.2&            13.19\\\hline
 \end{tabular}
\end{center}
\end{minipage}	

\begin{minipage}{\linewidth}
        \begin{center}
          \begin{tabular}{ c c c c c c c }
Cloudlet ID & Status & Data center ID & VM ID & Time & Start Time & Finish Time \\\hline
54&            SUCCESS&                 3&                              6&            12.99&        0.2&                  13.19 \\\hline   
 66&            SUCCESS&                 3                             & 6&            12.99&        0.2&                 13.19\\\hline 
 78&            SUCCESS&                 3&                              6&            12.99&        0.2&                  13.19\\\hline 
 90&            SUCCESS&                 3&                            6&            12.99&        0.2&                 13.19\\\hline 
 102&           SUCCESS&                   3&                              6&            12.99&        0.2&                13.19\\\hline 
     114&        SUCCESS&                   3&                              6&            12.99&        0.2&                13.19\\\hline 
 126&        SUCCESS&                   3&                              6&            12.99&        0.2&                13.19\\\hline 
 138&        SUCCESS&                   3&                              6&            12.99&        0.2&                13.19\\\hline 
 150&        SUCCESS&                   3&                              6&            12.99&        0.2&                13.19\\\hline 
7&            SUCCESS&                   3&                              7&            12.99&         0.2&               13.19\\\hline
    19&        SUCCESS&                     3&                             7&            12.99&         0.2&               13.19\\\hline
31&        SUCCESS&                      3&                            7&            12.99&        0.2&                13.19\\\hline
    43&        SUCCESS&                      3&                            7&            12.99&        0.2&                13.19\\\hline 
55&        SUCCESS&                      3&                            7&            12.99&        0.2&                13.19\\\hline
    67&       SUCCESS&                      3&                            7&            12.99&        0.2&                13.19\\\hline
79&        SUCCESS&                      3&                           7&            12.99&        0.2&                 13.19\\\hline
91&        SUCCESS&                      3&                           7&            12.99&        0.2&                 13.19\\\hline
103&        SUCCESS&                    3&                           7&            12.99&        0.2&                 13.19\\\hline
    115&        SUCCESS&                     3& 7&            12.99&        0.2&                13.19\\\hline
    127&        SUCCESS&                     3&                          7&            12.99&        0.2&                13.19\\\hline
       139&        SUCCESS&                      3&                        7&              12.99&        0.2&               13.19 \\\hline
   151&        SUCCESS&                      3&                       7&                12.99&        0.2&             13.19 \\\hline
       8&        SUCCESS&                           3&                      8&                 12.99&        0.2&             13.19 \\\hline
        20&        SUCCESS&                         3&                      8&                 12.99&        0.2&             13.19\\\hline
        32&        SUCCESS&                         3&                      8&                 12.99&         0.2&              13.19\\\hline
       44&        SUCCESS&                         3&                       8&                 12.99&        0.2&              13.19 \\\hline
       56&        SUCCESS&                         3&                       8&                 12.99&        0.2&              13.19 \\\hline
    68&        SUCCES&                         3&                       8&                12.99&         0.2&              13.19\\\hline
80&        SUCCESS&                         3&                       8& 12.99&        0.2&              13.19\\\hline
    92&        SUCCESS&                         3&                       8&                 12.99&        0.2&              13.19\\\hline
    104&        SUCCESS&                       3&                       8&                12.99&        0.2&              13.19\\\hline
    116&        SUCCESS&                       3&                       8&                12.99&        0.2&              13.19\\\hline
128&        SUCCESS&                       3&                       8&                12.99&        0.2&              13.19\\\hline
    140&        SUCCESS&                       3&                       8&                12.99&        0.2&             13.19\\\hline
152&        SUCCESS&                       3&                       8&                12.99&        0.2&             13.19\\\hline
10&        SUCCESS&                         3&                       10              & 12.99&        0.2&            13.19\\\hline
22&        SUCCESS&                         3&                       10&               12.99&        0.2&            13.19\\\hline
34&               SUCCESS&                        3&                      10&            12.99&        0.2&            13.19\\\hline
\end{tabular}
\end{center}
\end{minipage}	

\begin{minipage}{\linewidth}
        \begin{center}
          \begin{tabular}{ c c c c c c c }
Cloudlet ID & Status & Data center ID & VM ID & Time & Start Time & Finish Time \\\hline

    46&              SUCCESS&                        3&                      10&            12.99&        0.2&            13.19\\\hline

 58&             SUCCESS&                        3&                        10&            12.99&        0.2&            13.19\\\hline
 70&              SUCCESS&                       3                       & 10&            12.99&        0.2&            13.19\\\hline
 82&                  SUCCESS&                      3&                         10&            12.99&        0.2&            13.19\\\hline
 94&              SUCCESS&                     3&                         10&            12.99&        0.2&            13.19\\\hline
 106&            SUCCESS&                     3&                         10&            12.99&        0.2&            13.19\\\hline   
     118&            SUCCESS&                      3&                        10&            12.99&        0.2&            13.19\\\hline
     130&            SUCCESS&                      3&                        10&            12.99&        0.2&            13.19\\\hline
 142&            SUCCESS&                      3&                        10&            12.99&        0.2&            13.19\\\hline
 154&            SUCCESS&                      3&                        10&            12.99&        0.2&            13.19\\\hline
9&                SUCCESS&                      3&                          9&            12.99&        0.2&            13.19 \\\hline
    21&             SUCCESS&                      3&                           9&            12.99&        0.2&            13.19 \\\hline
 33&             SUCCESS&                      3&                           9&            12.99&        0.2&            13.19\\\hline
 45&             SUCCESS&                      3&                           9&            12.99&        0.2&            13.19\\\hline
    57&            SUCCESS&                        3&                          9&            12.99&        0.2&            13.19\\\hline
69&            SUCCESS&                        3&                          9&            12.99&        0.2&            13.19\\\hline
81&            SUCCESS&                        3&                          9&            12.99&        0.2&            13.19\\\hline
    93&            SUCCESS&                        3&                          9&            12.99&        0.2&            13.19\\\hline
    105&          SUCCESS&                        3&                          9&            12.99&        0.2&            13.19\\\hline
    117&          SUCCESS&                        3&                          9&            12.99&        0.2&            13.19\\\hline
129&          SUCCESS&                        3&                          9&            12.99&        0.2&            13.19\\\hline
141&          SUCCESS&                        3&                          9&            12.99&        0.2&            13.19\\\hline
153&         SUCCESS&                         3&                          9&            12.99&        0.2&            13.19\\\hline
11&           SUCCESS&                          3&                        11&            12.99&        0.2&            13.19\\\hline
23&           SUCCESS&                          3&                        11&            12.99&        0.2&            13.19\\\hline 
     35&          SUCCESS&                           3&                        11&            12.99&        0.2&            13.19\\\hline
    47&          SUCCESS&                           3&                        11&            12.99&        0.2&            13.19 \\\hline
 59&          SUCCESS&                           3&                        11&            12.99&        0.2&            13.19\\\hline
 71&          SUCCESS&                          3&                         11&            12.99&        0.2&            13.19\\\hline
83&          SUCCESS&                          3&                         11&            12.99&        0.2&            13.19\\\hline
95&          SUCCESS&                           3&                        11&            12.99&        0.2&            13.19\\\hline
107&        SUCCESS&                           3&                        11&            12.99&        0.2&            13.19\\\hline
    119&        SUCCESS&                           3&                        11&            12.99&        0.2&            13.19\\\hline
    131&        SUCCESS&                           3&                        11&            12.99&        0.2&            13.19\\\hline
 143&        SUCCESS&                           3&                         11&            12.99&        0.2&            13.19\\\hline
155&        SUCCESS&                           3&                         11&            12.99&        0.2&            13.19\\\hline
\end{tabular}
\end{center}
\end{minipage}

\begin{minipage}{\linewidth}
        \begin{center}
          \begin{tabular}{ c c c c c c c }
Cloudlet ID & Status & Data center ID & VM ID & Time & Start Time & Finish Time \\\hline
      0&                 SUCCESS&                 2&                               0&            14&            0.2&         14.2\\\hline
    12&                 SUCCESS&                  2&                               0&            14&           0.2&         14.2\\\hline
    24&                 SUCCESS&                   2&                              0&            14&          0.2&          14.2\\\hline
36&                SUCCESS&                   2&                               0&            14&         0.2&           14.2\\\hline
48&               SUCCESS&                   2&                                0&            14&        0.2&            14.2\\\hline
    60&             SUCCESS&                   2&                                 0&            14&        0.2&            14.2\\\hline
72&             SUCCESS&                   2&                                 0&            14&        0.2&            14.2\\\hline
84&             SUCCESS&                   2                                & 0&            14&        0.2&            14.2\\\hline
96&              SUCCESS&                  2                               & 0&            14&        0.2&            14.2\\\hline
    108&            SUCCESS&                  2&                                0&            14&        0.2&            14.2\\\hline
    120&            SUCCESS&                  2&                                 0&            14&        0.2&            14.2\\\hline
132&             SUCCESS&                 2&                                0&            14&        0.2&            14.2\\\hline
144&             SUCCESS&                 2&                                0&            14&        0.2&            14.2\\\hline
    156&             SUCCESS&                 2&                               0&            14&        0.2&            14.2\\\hline
1&                  SUCCESS&                2&                               1&            14&        0.2&            14.2\\\hline
13&                SUCCESS&                2&                               1&            14&        0.2&            14.2\\\hline    
 25&                SUCCESS&                2&                               1&            14&        0.2&            14.2\\\hline  
    37&               SUCCESS&                 2&                                1&            14&        0.2&            14.2 \\\hline  
49&               SUCCESS&                 2&                                1&            14&        0.2&            14.2 \\\hline  
 61&               SUCCESS&                2&                                1&            14&        0.2&            14.2\\\hline  
 73&               SUCCESS&               2&                                1&            14&        0.2&            14.2\\\hline  
 85&               SUCCESS&               2&                                1&            14&        0.2&            14.2\\\hline  
 97&               SUCCESS&               2& 1&            14&        0.2&            14.2\\\hline  
 109&             SUCCESS&               2&                                1&            14&        0.2&            14.2\\\hline  
 121&             SUCCESS&               2&                                1&            14&        0.2&            14.2\\\hline  
 133&             SUCCESS&               2&                                1&            14&        0.2&            14.2\\\hline  
 145&             SUCCESS&               2&                                1&            14&        0.2&            14.2\\\hline  
 157&             SUCCESS&               2&                                1&            14&        0.2&            14.2\\\hline  
2&                  SUCCESS&               2&                               2&            14&        0.2&            14.2 \\\hline     
 14&                SUCCESS&               2&                              2&            14&        0.2&            14.2\\\hline 
 26&                SUCCESS&               2&                               2&            14&        0.2&            14.2\\\hline 
     38&                SUCCESS&              2&                                2&            14&        0.2&            14.2\\\hline 
     50&                SUCCESS&              2&                                2&            14&        0.2&            14.2\\\hline 
     62&                SUCCESS&              2&                                2&            14&        0.2&            14.2\\\hline    
 74&                SUCCESS&              2&                                 2&            14&        0.2&            14.2\\\hline
 86&                SUCCESS&              2&                                 2&            14&        0.2&            14.2\\\hline
 98&                SUCCESS&              2&                                 2&            14&        0.2&            14.2\\\hline   
  \end{tabular}
\end{center}
\end{minipage}	  

\begin{minipage}{\linewidth}
        \begin{center}
          \begin{tabular}{ c c c c c c c }
Cloudlet ID & Status & Data center ID & VM ID & Time & Start Time & Finish Time \\\hline
   110&             SUCCESS&            2&                         2&                14&          0.2&            14.2\\\hline
    122&             SUCCESS&           2&                         2&                14&          0.2&            14.2\\\hline
    134&             SUCCESS&           2&                         2&                14&           0.2&            14.2\\\hline
146&             SUCCESS&          2&                        2&                  14&            0.2&            14.2\\\hline
158&          SUCCESS&             2&                        2&                   14&            0.2&            14.2\\\hline
3&               SUCCESS&         2&                          3&                    14&            0.2&            14.2\\\hline
15&               SUCCESS&        2&                          3&                    14&            0.2&            14.2\\\hline
27&               SUCCESS&        2&                          3&                     14&            0.2&            14.2\\\hline
    39&               SUCCESS&        2&                          3&                     14&            0.2&            14.2\\\hline
51&               SUCCESS&        2&                          3&                     14&            0.2&            14.2\\\hline
63&               SUCCESS&        2&                          3&                     14&            0.2&            14.2\\\hline
75&               SUCCESS&        2&                          3&                     14&            0.2&            14.2\\\hline
87&               SUCCESS&        2&                          3&                      14&            0.2&            14.2\\\hline
99&               SUCCESS&        2&                          3&                       14&            0.2&            14.2\\\hline
111&             SUCCESS&        2&                          3&                        14&            0.2&            14.2\\\hline
123&             SUCCESS&        2&                           3&                        14&           0.2&            14.2\\\hline
    135&             SUCCESS&        2&                           3&                        14&           0.2&            14.2\\\hline
    147&             SUCCESS&        2&                            3&                       14&            0.2&            14.2\\\hline
159&             SUCCESS&        2&                            3&                        14&           0.2&            14.2\\\hline
\end{tabular}
\end{center}
\end{minipage}
We observed that the time is  four to five times higher when compared to the previous run and the number of successful cloudlet performances has improved dramatically.
\section{Conclusion}
We found that CloudSim supports a large scale simulation environment with little overhead and it exposes powerful features that could easily be extended for modeling custom Cloud environments. We described the CloudSim architecture and then investigated the new models and techniques in CloudSim.                                                                                                                                                                                                                                                                                     We calculated two values in our research, the completion rate and the average execution time  in touch CloudSim.        

\end{document}